\documentclass[prc,twocolumn,showpacs,superscriptaddress,preprintnumbers,amsmath,amssymb]{revtex4-1}
\usepackage{bm}
\usepackage{dcolumn}
\usepackage{multirow}
\usepackage{ifpdf}
\usepackage{url}
\usepackage[utf8]{inputenc}

\ifpdf
  \usepackage{graphicx}     
  \usepackage{hyperref}
\else     
  \usepackage[dvipdfmx]{graphicx}     
  \usepackage[dvipdfmx]{hyperref}
\fi

\newcommand\pythia{\textsc{Pythia}}

\usepackage[normalem]{ulem}  
\usepackage[dvips]{color} 

\renewcommand\sout{\bgroup \color{red} \ULdepth=-.5ex \ULset}

\hypersetup{
colorlinks=true,
linkcolor=blue,
citecolor=blue,
urlcolor=blue
}

\begin{document}

\title{
Enhancement of strange baryons\\
 in high-multiplicity  
proton--proton and proton--nucleus collisions
}

\author{Yuuka Kanakubo}
\email{y-kanakubo-75t@eagle.sophia.ac.jp}
\affiliation{%
Department of Physics, Sophia University, Tokyo 102-8554, Japan
}

\author{Michito Okai}
\email{michito0605@eagle.sophia.ac.jp}
\affiliation{%
Department of Physics, Sophia University, Tokyo 102-8554, Japan
}
\author{Yasuki Tachibana}
\email{yasuki.tachibana@wayne.edu}
\affiliation{
Department of Physics and Astronomy, Wayne State University, Detroit, Michigan 48201, USA 
}
\affiliation{%
Department of Physics, Sophia University, Tokyo 102-8554, Japan
}
\author{Tetsufumi Hirano}
\email{hirano@sophia.ac.jp}
\affiliation{%
Department of Physics, Sophia University, Tokyo 102-8554, Japan
}

\date{\today}

\begin{abstract}
We investigate the enhancement of yields of strange
and multi-strange 
 baryons
in proton--proton (p+p), 
proton--lead (p+Pb)
and lead--lead (Pb+Pb) 
collisions at the Large Hadron Collider (LHC) energies
from a dynamical core--corona initialization model. 
We first generate partons just after the collisions by using event generators.
These partons dynamically generate the quark gluon plasma (QGP)
fluids through the source terms in the hydrodynamic equations.
According to the core--corona picture, this process 
tends to happen where the density of generated partons is high
and their transverse momentum is low. 
Some partons do not fully participate in this process when they are in dilute regions or their transverse momentum is high and subsequently fragment into hadrons through string fragmentation.
In this framework, the final hadrons come from either chemically equilibrated fluids as in the conventional
hydrodynamic models or string fragmentation.
We calculate the ratio of strange baryons to charged pions as 
a function of multiplicity
and find that it monotonically increases up to $dN_{\mathrm{ch}}/d\eta \sim 100$ and then saturates above.
This suggests that the QGP fluids are \textit{partly} created and that their fraction increases with multiplicity
in p+p and p+Pb collisions at LHC energies.
\end{abstract}


\maketitle

\textit{Introduction.}---
High-energy heavy-ion collision experiments are performed 
at the Relativistic heavy-ion Collider (RHIC), Brookhaven National Laboratory,
and the Large Hadron Collider (LHC), CERN,
 to further understanding of the properties of deconfined nuclear matter,
the quark gluon plasma (QGP) \cite{Yagi:2005yb}.
A vast body of the experimental data have been accumulated and theoretical analysis of them
elucidates that the QGP behaves almost like a perfect fluid
\cite{Heinz:2001xi,sQGP1,sQGP2,sQGP3,Hirano:2005wx}. 

Comparisons of data from heavy-ion collision experiments with 
those from control experiments such as proton--proton, proton--nucleus and deuteron--nucleus
collisions could bring deeper insights into the properties of the QGP.
However, high-multiplicity events in these small colliding systems
exhibit some collective behaviors, 
which can be interpreted as creation of QGP fluids 
(for a review, see, e.g., Ref.~\cite{Dusling:2015gta}).
In addition, 
enhanced production of multi-strange hadrons relative to charged pions 
has been measured in high-multiplicity small colliding systems \cite{ALICE:2017jyt}. 
Strangeness enhancement 
was proposed as a signature of QGP formation,
\cite{Rafelski:1982pu,Koch:1982ij,Koch:1986ud}
and has been observed in high-energy heavy-ion collisions
\cite{Andersen:1999ym,Afanasiev:2002he,Antinori:2004ee,Abelev:2007xp,ABELEV:2013zaa}.
The ratio of yields of multi-strange hadrons to those of charged pions
monotonically increases with charged hadron
multiplicity at mid-rapidity, $dN_{\mathrm{ch}}/d\eta$, 
and saturates above $dN_{\mathrm{ch}}/d\eta \sim 100$
regardless of the size or collision energy of the systems \cite{ALICE:2017jyt}.
In the low-multiplicity limit, the ratio can be described by 
string fragmentation \cite{Sjostrand:2007gs}. 
On the other hand, 
the saturated value of the ratio mainly in Pb+Pb collisions
can be interpreted 
as hadron production from chemically equilibrated hadronic matter with (anti-)strangeness
through statistical model analysis.
(See, \textit{e.g.}, Ref.~\cite{Andronic:2017pug}).
Thus the increasing behavior of the ratio suggests 
a continuous change of the hadron production mechanism 
from fragmentation dominance to the chemically-equilibrated-matter dominance.

In this Letter,
we develop a dynamical core--corona initialization model
to investigate 
the production of QGP fluids in p+p, p+Pb and Pb+Pb collisions 
at LHC energies and show that 
strangeness enhancement is controlled 
by multiplicity 
as implied by the ALICE data \cite{ALICE:2017jyt}
, rather than the size of the colliding systems.
This model describes the dynamics of  
gradually forming QGP fluids 
as the density of the produced partons increases
according to the ``core--corona'' picture \cite{Werner:2007bf,Aichelin:2008mi,Becattini:2008ya,Steinheimer:2011mp,Pierog:2013ria,Petrovici:2017izo,Werner:2018SQM,
Akamatsu:2018olk,Bozek:2005eu,Bozek:2008zw}: 
When a parton produced in the very early stage propagates through 
regions occupied by many other partons (the ``core''), 
the parton deposits its energy and momentum
due to the strong interactions among them
 and gives rise to locally equilibrated fluids. 
In contrast, a parton propagating through dilute areas (the ``corona'') does
not take part in the formation of the fluids
and undergoes vacuum fragmentation. 

In the following, we first formulate
 the dynamical core--corona initialization model and
 then perform numerical simulations in various colliding systems such as
p+p at $\sqrt{s_{NN}}=7$ TeV, 
p+Pb at $5.02$ TeV, and Pb+Pb at $2.76$ TeV.
We estimate 
to what extent energy and momentum of the particles created in the initial collisions 
are converted into the medium fluid. 
Finally we study the multiplicity dependence of the ratios of (multi-)strange hadron yields to 
charged pion yields. 

We use the natural unit, $\hbar$ = $c$ = $k_{B}$ = $1$, and the Minkowski metric,
$g^{\mu \nu} = \mathrm{diag}(1, -1, -1, -1)$, 
throughout this paper. 
We also use 
the Milne coordinates,
$\left(\tau, x, y, \eta_{s} \right)=\left(\tau, \vec{x}_\perp, \eta_{s} \right)$, 
where 
$\tau=\sqrt{t^2-z^2}$ is the proper time 
and 
$\eta_{s}=\left(1/2\right)\ln\left[\left(t+z\right)/\left(t-z\right)\right]$ is 
the spacetime rapidity.

\textit{Model.}---
In our framework, all matter created in high-energy proton--proton, proton--nucleus,
and nucleus--nucleus collisions 
originates from partons produced in the primary collisions.
Here we employ an event generator, \pythia\ 8.230 \cite{Sjostrand:2007gs}, for the production of partons. \pythia\ is a general- purpose event generator to capture a global feature of elementary particle reactions and has been utilized widely in the community. It is noted that heavy-ion reactions at high energies become available  from this version. The particle production model for heavy-ion reactions in \pythia\ is based on the improved version of \textsc{Fritiof} model \cite{ANDERSSON1987289, Bierlich:2016smv}. The partons are generated from \pythia\ by switching on the parton vertex information and switching off the hadronization process. After the production, the partons propagate along their eikonal path, 
$\bm{x}_i(t)=({\bm{p}^{\rm init}_i}/E^{\rm init}_i)t+\bm{x}^{\rm ver}_i$, 
where $\bm{x}_i$ is the position of the $i$-th parton at time $t$.
$E^{\rm init}_i$, $\bm{p}_i^{\rm init}$ and $\bm{x}^{\rm ver}_i$ 
are the initial energy, momentum and creation position in the parton vertex information
of the $i$-th parton, respectively, obtained from \pythia.

Then these partons deposit their energy and momentum into vacuum or fluids during their propagations. 
We model their energy-momentum deposition rate of the form, 
\begin{eqnarray}
\label{eq:hydrorate}
\frac{d p_i^\mu}{dt}(t) & = & -a_0  \frac{\rho_i (\bm{x}_i(t))}{{p_{T, i}}^2(t)} p_i^\mu (t) ,  \\ 
\label{eq:densitydistribution}
\rho_i(\bm{x}) &  = &\sum_{j\neq i} G(\bm{x} - \bm{x}_j(t)),
\end{eqnarray}
where $p_i^\mu$ is the four-momentum of the $i$ th parton
and the summation is taken over all partons in an event.
In the actual calculations, we solve Eq.~(\ref{eq:hydrorate})
 in the Milne coordinates.
Then, a smearing Gaussian function $G$ is
\begin{eqnarray}
\label{eq:gaussian}
&&G(\bm{x}-\bm{x}_i(t))d^3x \nonumber \\
& \rightarrow & \frac{1}{2\pi\sigma_\perp^2}\exp\left\{-\frac{[\vec{x}_\perp-\vec{x}_{\perp,i}(\tau)]^2}{2\sigma_\perp^2} \right\}\nonumber \\
& \times & \frac{1}{\sqrt{2\pi \tau^2 \sigma_{\eta_s}^2}}\exp\left\{-\frac{[\eta_s-\eta_{s,i}(\tau)]^2}{2\sigma_{\eta_s}^2} \right\} d^2\vec{x}_\perp \tau d\eta_{s}.
\end{eqnarray}
Here we assume that 
the fluidization rate is proportional to the spatial density of the partons
surrounding the $i$ th parton, $\rho_i(\bm{x})$, in order to apply the core--corona picture.
Note here that $\rho_{i}$ does not contain contribution from fluids.
In regions with high parton density (core), 
fluids are supposed to be created.
On the other hand, 
in low parton density regions (corona), 
fluids are not likely to be created. 
The factor $p_{T}^{\ -2}$ in Eq.~(\ref{eq:hydrorate}), 
which has the same dimension as the cross section, 
makes the soft partons tend to become fluids.
The dimensionless factor, $a_0$, is 
a parameter to control the overall strength of the fluidization process. 

The dynamical initialization of the hydrodynamic fields 
can be described by 
relativistic hydrodynamic equations with source terms \cite{Akamatsu:2018olk,Okai:2017ofp, Shen:2017bsr}, 
\begin{equation}
\label{eq:qgp+jet}
\partial_\mu T_{\rm fluid}^{\mu \nu} \!\left(x\right)= J^\nu\!\!\left(x\right).
\end{equation}
Here $T_{\rm fluid}^{\mu \nu}$ is the energy-momentum tensor of the fluids
and $J^{\nu}$  is the source term.
Since the matter produced at LHC energies is almost baryon free around mid-rapidity, 
we do not solve the continuity equation for the baryon number conservation. 
Energy-momentum tensor is modeled as the ideal one,
 $T_{\rm fluid}^{\mu \nu}=(e+P)u^{\mu}u^{\nu}-Pg^{\mu \nu}$,  
where $e$ is the energy density, 
$P$ is the pressure, 
and $u^{\mu}$ is the four-velocity of the fluid. 
The initial conditions of the hydrodynamic fields 
at the formation time of the produced partons, $\tau_{00}$,
are set to 
$T_{\rm fluid}^{\mu \nu}(\tau=\tau_{00}, \vec{x}_\perp, \eta_{s}) = 0$. 
The source term $J^{\nu}$ transfers the energy and momentum deposited from partons 
calculated as in Eq.~(\ref{eq:hydrorate}) into the hydrodynamic fields. 
Assuming that the deposited energy and momentum are instantaneously equilibrated around the partons, 
we employ a simple form of the source term with a Gaussian smearing, 
\begin{equation}
\label{eq:source_j}
J^\mu(x)  
=  -\sum_{i}  \frac{dp_i^\mu}{dt} G(\bm{x}-\bm{x}_i(t)). 
\end{equation}
From  $\tau_{00}$ to  $\tau_{0}$ (the hydrodynamic initial time in a conventional sense),
we numerically solve Eqs.~(\ref{eq:hydrorate}) and (\ref{eq:qgp+jet})  
simultaneously 
in $(3+1)$-dimensional spacetime to initialize the hydrodynamic fields.  
Motivated by the fact that the experimental results of the yield ratio between multi-strange hadrons and charged pions in Pb+Pb collisions are well reproduced by
a statistical model \cite{Andronic:2017pug}, 
we assume in this study that the fluids consist of chemically equilibrated matter with (anti-)strangeness.
Following this assumption, we employ an equation of state with (2+1) flavors from a lattice QCD result \cite{Borsanyi:2013cga}. 

At each time step, 
we trace a color flow of partons provided by \pythia\ to form a color singlet string
 and calculate its mass. 
When the string mass becomes lower than its threshold 
for undergoing string fragmentation as given in \pythia, 
all the partons in that string are assumed to be completely fluidized 
and all their energy and momentum are put into fluids 
through the source terms, Eq. (\ref{eq:source_j}). 
It should be noted that the sums of energy and of the momentum in the total system 
(fluids and partons)
are conserved in this framework all the way through the dynamical initialization. 

After $\tau_0$, 
the dynamics of the medium is the same as that in a conventional hydrodynamic approach: 
We solve Eq.~(\ref{eq:qgp+jet}) without source terms
until the maximum temperature goes below a fixed decoupling temperature, $T=T_{\mathrm{dec}}$.
In this study
 we neglect further energy and momentum loss 
of partons traversing after $\tau_{0}$ for simplicity,
as already discussed in Ref.~\cite{Okai:2017ofp}.

To obtain yields of hadrons directly emitted from the decoupling
hypersurface of the chemically equilibrated fluids, 
we use the Cooper--Frye formula \cite{Cooper:1974mv},
\begin{eqnarray}
N_i
&=&
\frac{g_i}{(2\pi)^3} \int \frac{d^3 p}{p^0}
\int_{\Sigma}
\frac{p^{\mu} d\sigma_{\mu}(x)}{\exp\left[ {p^{\mu}u_{\mu}\left(x\right)}/{T_{\mathrm{dec}}}\right]\mp_\mathrm{BF}1}, 
\label{eq:C-F}
\end{eqnarray}
where 
$g_i$ 
is the degeneracy, 
$\mp_{\rm BF}$ corresponds to 
Bose or Fermi statistics for hadron species $i$, 
$\Sigma$ is the decoupling hypersurface of $T(x) = T_{\mathrm{dec}}$, 
and $d\sigma^\mu$ is the normal vector of its element. 
Since we assume baryon free matter, 
the chemical potential for the baryon number or the strangeness does not appear in Eq.~(\ref{eq:C-F}).
Thus the net strangeness is neutral in the calculations.
For the contribution from resonance decays, 
we simply correct the direct yields by multiplying by a factor which is the ratio of the total yields to the contribution from directly produced hadrons estimated from Fig. 2 of Ref.~\cite{Andronic:2017pug}.
For a more rigorous treatment, 
the effects of hadronic rescatterings and decays of resonances
could be estimated through hadronic transport models 
as studied in Ref.~\cite{Takeuchi:2015ana}. 
We leave such an analysis as a future study. 

We push the surviving partons back into \pythia\,
with their reduced energy and momentum after completing the dynamical initialization
at $\tau_0$. 
Then these partons 
form a string with a mass above its threshold 
and are forced to hadronize through string fragmentation. 
Although resonances decay into stable particles in the default
setting in \pythia,
we switch off the decays of the neutral pions
and the strange baryons that are stable against strong decays  
(except for $\Sigma^0 \rightarrow \Lambda + \gamma$), 
so that we obtain their yields efficiently 
instead of performing mass reconstruction.

Thus the final hadrons in this study 
come from either chemically equilibrated fluids or string fragmentation.

\textit{Results.}---
The parameters in this model are summarized here.
The formation time and hydrodynamic initial time are $\tau_{00} = 0.1$ fm and $\tau_{0} = 0.6$ fm, respectively. 
To reproduce the ALICE data reasonably well,
we set the parameter to control the overall magnitude of fluidization rate
to be $a_0 = 100$.
The width parameters 
in the smearing Gaussian functions are $\sigma_\perp = 0.5$ fm
 and $\sigma_{\eta_{s}} = 0.5$.
For the moment, these parameters just regularize the numerical simulations to avoid spiky distributions. It would , however, be highly related to the coarse-graining processes in hydrodynamics, which is of particular interest for future study.
The decoupling temperature that is identified with the chemical freezeout temperature
in this approach is 
$T_{\rm dec}=160\,{\rm MeV}$. 
The correction factors that are multiplied with the direct hadron yields from the fluids obtained from Eq.~(\ref{eq:C-F}) 
to include the resonance decay contributions are 
$c_\pi = 3.2$ for pions,  $c_{\Lambda} = 4.7$ for lambdas and $c_{\Xi} = 1.7$ for cascades \cite{Andronic:2017pug}.
In what follows, 
the multiplicity at mid-rapidity, $dN_{\mathrm{ch}}/d\eta$ in $|\eta|<0.5$, 
in p+p, p+Pb and Pb+Pb collisions is obtained from the default calculations in \pythia\ 8.230.

We first analyze a fraction of the fluidized energy in the transverse plane at mid-rapidity 
as a function of multiplicity, $dN_{\mathrm{ch}}/d\eta$,
to estimate to what extent a QGP fluid is formed in an event.
The fluidized energy in the transverse plane with the core--corona picture is
\begin{equation}
\label{eq:energy_fluid}
\frac{dE_{\mathrm{core}}}{d\eta_{s}}  =  \int_{\tau_{00}}^{\tau_{0}} d\tau \int   d^2\vec{x}_\perp
 \tau J^\tau(\tau, \vec{x}_\perp,  \eta_{s}).
\end{equation}
One can also estimate the total energy in the transverse plane, 
$dE_{\mathrm{tot}}/d\eta_{s}$,
in a similar way
that all the initial partons
are forced to be fluidized in the first time step in Eq.~(\ref{eq:energy_fluid}).
Thus we obtain a fraction of the fluidized energy with the core--corona picture
as
\begin{equation}
\label{eq:energy_parton}
R= \frac{dE_{\mathrm{core}}/d\eta_s}{dE_{\mathrm{tot}}/d\eta_{s}}.
\end{equation}
\begin{figure}[htbp]
\begin{center}
\includegraphics[bb=0 0 340 369,width=0.50\textwidth]{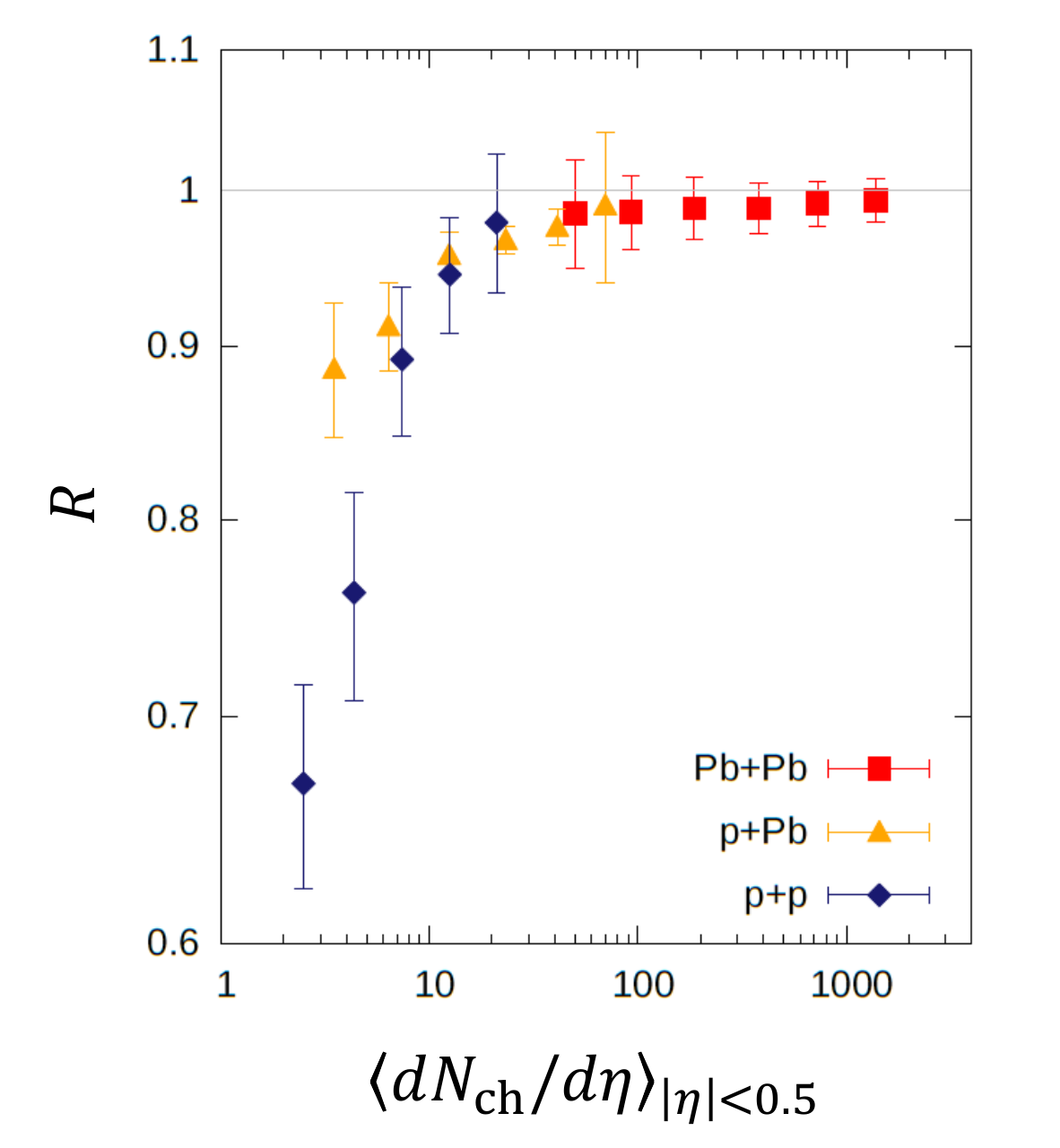}
\end{center}
\caption{(Color online)
Fraction of the fluidized energy to the total energy
at $\eta_{s} = 0$ 
as a function of multiplicity at mid-rapidity, $dN_{\mathrm{ch}}/d\eta$ ($|\eta|<0.5$). 
The center-of-mass collision energy per nucleon pair, $\sqrt{s_{NN}}$, is 
7 TeV in p+p (diamonds), 5.02 TeV in p+Pb (triangles) and 2.76 TeV in Pb+Pb 
(squares) collisions.
}
\label{fig:fluid-energy-fraction}
\end{figure} 

Figure \ref{fig:fluid-energy-fraction} shows the fractions of the fluidized
energy at $\eta_{s} = 0$ as functions of multiplicity at mid-rapidity, $dN_{\mathrm{ch}}/d\eta$ ($|\eta |<0.5$),
in p+p, p+Pb and Pb+Pb collisions at the LHC energies.
It should be noted that the hydrodynamic simulations are performed
in the center-of-mass frame and that the particle yields are counted in the
laboratory frame. Hence there exists a rapidity shift, $\Delta \eta_{s} = 0.47$, 
between the laboratory frame and the center-of-mass frame in p+Pb collisions
at $\sqrt{s_{NN}}=5.02$ TeV.
The fraction of the fluidized energy increases with multiplicity
and saturates above $dN_{\mathrm{ch}}/d\eta \sim 50$-$100$ regardless of the collision 
systems or energy. 
This clearly demonstrates that the core--corona picture is properly incorporated
in the dynamical initialization model through Eqs.~(\ref{eq:hydrorate}) and (\ref{eq:source_j}).

\begin{figure*}[htbp]
\begin{center}
\includegraphics[bb=0 0 360 540,width=0.45\textwidth]{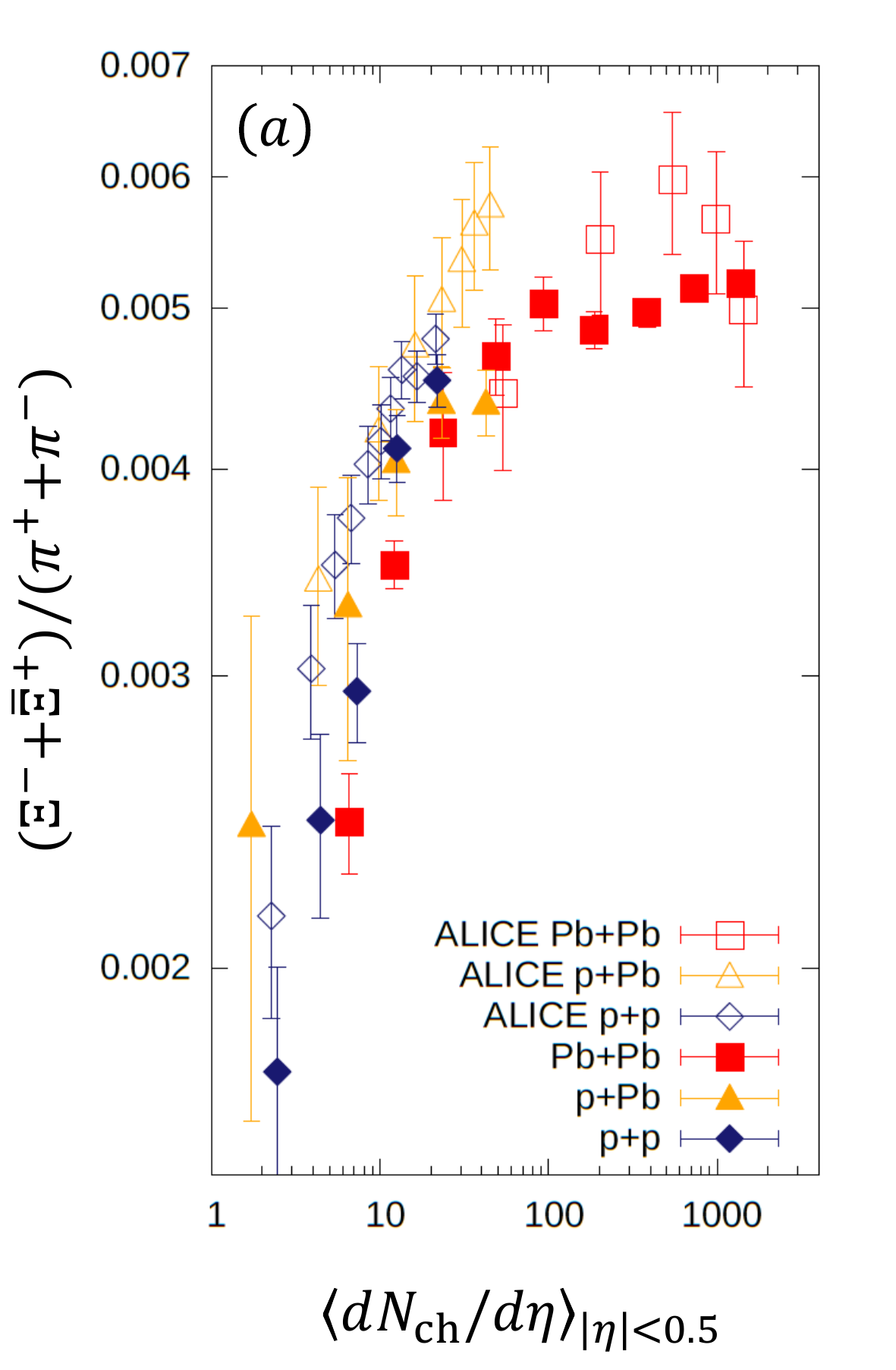}
\hspace{25pt}
\includegraphics[bb=0 0 360 540,width=0.45\textwidth]{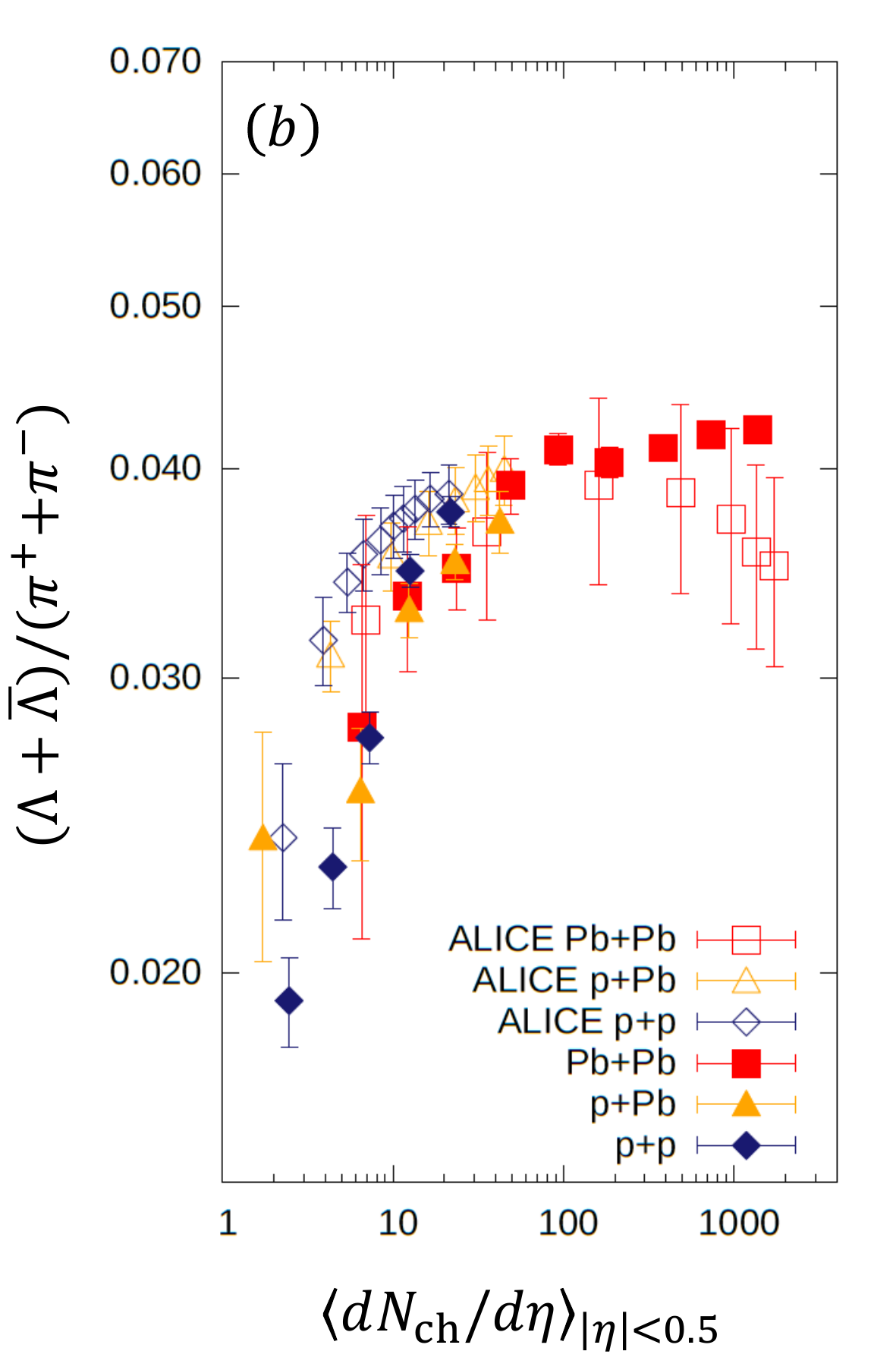}
\end{center}
\caption{(Color online)
Ratio of yields of (a) cascades ($\Xi^{-}$ and $\bar{\Xi}^{+}$) 
and (b) lambdas ($\Lambda$ and $\bar{\Lambda}$)
to
the ones of charged pions ($\pi^-$ and $\pi^{+}$)
as a function of multiplicity at mid-rapidity, $dN_{\mathrm{ch}}/d\eta$,
in p+p (diamonds), p+Pb (triangles)
and Pb+Pb (squares) collisions at the LHC energies.
The center-of-mass collision energy per nucleon pair is, $\sqrt{s_{NN}}=7$
TeV, $5.02$ TeV and $2.76$ TeV in p+p,  p+Pb and Pb+Pb collisions, respectively.
Results from a dynamical core--corona initialization model (closed symbols) are
compared with  ALICE data (open symbols) \cite{ABELEV:2013zaa,
Abelev:2013haa,
Adam:2015vsf,
ALICE:2017jyt}.
}
\label{fig:s-ratio}
\end{figure*}

Figure \ref{fig:s-ratio} (a) shows 
the ratio of the yields of cascades ($\Xi^{-}$ and $\bar{\Xi}^{+}$) 
to those of charged pions ($\pi^-$ and $\pi^{+}$)
as a function of multiplicity in $|\eta| < 0.5$ in p+p (at $\sqrt{s}=7$ TeV),
p+Pb (at $\sqrt{s_{NN}} = 5.02$ TeV) and Pb+Pb (at $\sqrt{s_{NN}}=2.76$ TeV) collisions, 
compared to the experimental data from the ALICE Collaboration \cite{ABELEV:2013zaa,Adam:2015vsf,ALICE:2017jyt}. 
The yields of final hadrons in our results are 
the sum of the contribution from chemically equilibrated fluids 
and from string fragmentation.
When we calculate the particle yields from the fluid part via Eq.~(\ref{eq:C-F}), 
the value at $\eta_p =0$ is used where $\eta_p$ is the momentum rapidity.
The yields of pions and cascades obtained in this way
 are corrected by multiplying by resonance correction factors
$c_\pi$ and $c_\Xi$, respectively.
On the other hand, 
when we obtain the yields from string fragmentation via \pythia,
the rapidity region is taken to be $|\eta_p|<2.0$ to gain statistics
and divided by $\Delta \eta_p = 4.0$.
The error bars in these results are statistical ones and originate only from string fragmentations, while those in experimental data
are systematical ones.
The results with the dynamical core--corona initialization model
capture the tendency of the data, 
i.e., the monotonic increase up to $dN_{\mathrm{ch}}/d\eta \sim 100$ 
and the saturation above it. 
In the very low multiplicity events, $dN_{\mathrm{ch}}/d\eta \sim 1$, 
the ratio is close to the vacuum fragmentation limit, $N_{\Xi}/N_{\pi} \sim 0.002$, 
which is estimated from the calculation solely by \pythia\ with the Lund string fragmentation. 
As the multiplicity increases, 
the fraction of the QGP fluid formation increases regardless of colliding systems 
and the ratio reaches the limit of chemically equilibrated fluids, $N_{\Xi}/N_{\pi} \sim 0.005$, 
which can be estimated by the statistical models (e.g., Ref.~\cite{Andronic:2017pug}). 
This behavior is deduced quite naturally 
from a fraction of the fluidized energy shown in Fig.~\ref{fig:fluid-energy-fraction}.

Shown in Fig.~\ref{fig:s-ratio} (b) is 
the ratio of yields of lambdas ($\Lambda$ and $\bar{\Lambda}$) to
those of charged pions ($\pi^-$ and $\pi^{+}$) as a function of multiplicity, 
compared to the ALICE data \cite{Abelev:2013haa,ALICE:2017jyt}. 
Here the same behavior is seen as for cascades  in Fig.~\ref{fig:s-ratio} (a): 
a monotonic increase up to $dN_{\mathrm{ch}}/d\eta \sim 100$ 
and saturation above it. 
The ratio is close to the fragmentation limit $N_{\Lambda}/N_{\pi} \sim 0.025$ 
for the very low multiplicity events. 
Then it monotonically increases with the multiplicity, 
and saturates around $dN_{\mathrm{ch}}/d\eta = 100$ with 
the value, $N_{\Lambda}/N_{\pi} \sim 0.04$, 
estimated by the statistical models. 
The dynamical core--corona initialization model
demonstrates 
that the enhancement of lambdas is less prominent than that of cascades
as a function of multiplicity, depending on their strangeness quantum number.

\textit{Summary.}---
In this letter we have formulated a dynamical initialization model with the core--corona picture
to analyze hadron yields from chemically equilibrated fluids and
string fragmentation 
in high-energy proton--proton, proton--nucleus and nucleus--nucleus collisions.
From this model, we calculated the ratios of strange baryon yields to charged pion yields
as functions of multiplicity and compared them with the ALICE data.

In this model, 
all the matter is initialized from partons created 
in the primary collisions. 
These partons were generated from \pythia\ 8.230. 
They deposited their energy and momentum 
so as to create QGP fluids via the source terms of hydrodynamic equations 
during their propagation in the vacuum or fluids. 
The energy-momentum deposition rate was parametrized 
to capture the core--corona picture. 
The rate is higher in the denser region of the generated partons 
to more likely generate the QGP fluids. 
On the other hand, the QGP fluids are generated less in the more dilute region.
Partons with higher $p_{T}$ contribute less to this dynamical initialization process. 
After the dynamical initialization,
the fluid expands and cools down as it generates hadrons from the decoupling hypersurface.
Here the ratios of hadron yields from the chemically equilibrated fluids 
reflect the decoupling temperature regardless of multiplicity.
On the other hand, the partons surviving after the dynamical initialization
undergo string fragmentation in \pythia. 
The string fragmentation gives 
hadron yield ratios, being almost independent of multiplicity, different 
from those from the chemically equilibrated fluids. 
As a result, the ratios are between the value at the chemically equilibrated fluid limit and
that at the fragmentation limit according to their fraction. 
We found the ratio of $\Xi$ or $\Lambda$ to $\pi$ increases with multiplicity
from the fragmentation limit, reaches the chemically equilibrated fluid limit
at $dN_{\mathrm{ch}}/d\eta \sim 100$, and saturates above.
This strongly suggests that the QGP fluids are partly produced in high-multiplicity
p+p and p+Pb collisions although the contribution from fragmentation is still important.
It also suggests that the contribution from the QGP fluids is dominant in hadron yields in Pb+Pb collisions, if multiplicity at mid-rapidity exceeds $dN_{\mathrm{ch}}/d\eta \sim 100$. 
One may access more detailed information about 
the equilibration in the QGP fluid by studying the production of other particle species. 
For example, investigation of the ratio of $\phi$ mesons to pions enables us to discriminate between 
the effect from the core--corona picture and that from the canonical strangeness suppression \cite{Becattini:2008ya}. 
We will cover these analyses, together with collision energy and model-parameter dependences on the ratios, 
in a future publication \cite{KanakuboFull}.

In the dynamical core--corona initialization model employed in this study, 
soft and hard physics can be treated in a unified manner. 
According to the parametrization of $p_{T}$ dependence in the fluidization rate, 
relatively more partons with higher $p_{T}$ survive, 
which produces the dominant sources of final hadrons in high $p_{T}$ regions, 
while low-$p_{T}$ hadrons in the final state are dominated 
by the hydrodynamic component. 
Since soft and hard particles are treated together, 
the resulting spectra can cover the entire momentum region 
and correlations between soft and hard physics are encoded naturally in this framework. 
In this study we focused on 
the yield ratio between (multi-)strange baryons and charged pions. 
It would also be interesting to see flow observables, 
such as two-particle correlation functions and anisotropic flow parameters 
in azimuthal distributions. 
In particular, in 
moderate-multiplicity events,
the competition between the fragmentation component and the hydrodynamic component 
plays a significant role. 
We will report on this analysis elsewhere.

\section*{Acknowledgement}
Y.K. and T.H. thank K.~Murase for useful discussions. 
The work by Y.T. was supported in part by a special award from 
the Office of the Vice President of Research at Wayne State University 
and in part by the National Science Foundation (NSF) 
within the framework of the JETSCAPE collaboration under Award No. ACI-1550300.
The work by T.H. was supported by JSPS KAKENHI Grant Number JP17H02900.

\appendix
\bibliography{ref}

\begin{thebibliography}{38}%
\makeatletter
\providecommand \@ifxundefined [1]{%
 \@ifx{#1\undefined}
}%
\providecommand \@ifnum [1]{%
 \ifnum #1\expandafter \@firstoftwo
 \else \expandafter \@secondoftwo
 \fi
}%
\providecommand \@ifx [1]{%
 \ifx #1\expandafter \@firstoftwo
 \else \expandafter \@secondoftwo
 \fi
}%
\providecommand \natexlab [1]{#1}%
\providecommand \enquote  [1]{``#1''}%
\providecommand \bibnamefont  [1]{#1}%
\providecommand \bibfnamefont [1]{#1}%
\providecommand \citenamefont [1]{#1}%
\providecommand \href@noop [0]{\@secondoftwo}%
\providecommand \href [0]{\begingroup \@sanitize@url \@href}%
\providecommand \@href[1]{\@@startlink{#1}\@@href}%
\providecommand \@@href[1]{\endgroup#1\@@endlink}%
\providecommand \@sanitize@url [0]{\catcode `\\12\catcode `\$12\catcode
  `\&12\catcode `\#12\catcode `\^12\catcode `\_12\catcode `\%12\relax}%
\providecommand \@@startlink[1]{}%
\providecommand \@@endlink[0]{}%
\providecommand \url  [0]{\begingroup\@sanitize@url \@url }%
\providecommand \@url [1]{\endgroup\@href {#1}{\urlprefix }}%
\providecommand \urlprefix  [0]{URL }%
\providecommand \Eprint [0]{\href }%
\providecommand \doibase [0]{http://dx.doi.org/}%
\providecommand \selectlanguage [0]{\@gobble}%
\providecommand \bibinfo  [0]{\@secondoftwo}%
\providecommand \bibfield  [0]{\@secondoftwo}%
\providecommand \translation [1]{[#1]}%
\providecommand \BibitemOpen [0]{}%
\providecommand \bibitemStop [0]{}%
\providecommand \bibitemNoStop [0]{.\EOS\space}%
\providecommand \EOS [0]{\spacefactor3000\relax}%
\providecommand \BibitemShut  [1]{\csname bibitem#1\endcsname}%
\let\auto@bib@innerbib\@empty
\bibitem [{\citenamefont {Yagi}\ \emph {et~al.}(2005)\citenamefont {Yagi},
  \citenamefont {Hatsuda},\ and\ \citenamefont {Miake}}]{Yagi:2005yb}%
  \BibitemOpen
  \bibfield  {author} {\bibinfo {author} {\bibfnamefont {K.}~\bibnamefont
  {Yagi}}, \bibinfo {author} {\bibfnamefont {T.}~\bibnamefont {Hatsuda}}, \
  and\ \bibinfo {author} {\bibfnamefont {Y.}~\bibnamefont {Miake}},\
  }\href@noop {} {\bibfield  {journal} {\bibinfo  {journal} {Camb. Monogr.
  Part. Phys. Nucl. Phys. Cosmol.}\ }\textbf {\bibinfo {volume} {23}},\
  \bibinfo {pages} {1} (\bibinfo {year} {2005})}\BibitemShut {NoStop}%
\bibitem [{\citenamefont {Heinz}\ and\ \citenamefont
  {Kolb}(2002)}]{Heinz:2001xi}%
  \BibitemOpen
  \bibfield  {author} {\bibinfo {author} {\bibfnamefont {U.~W.}\ \bibnamefont
  {Heinz}}\ and\ \bibinfo {author} {\bibfnamefont {P.~F.}\ \bibnamefont
  {Kolb}},\ }\href {\doibase 10.1016/S0375-9474(02)00714-5} {\bibfield
  {journal} {\bibinfo  {journal} {Nucl. Phys.}\ }\textbf {\bibinfo {volume}
  {A702}},\ \bibinfo {pages} {269} (\bibinfo {year} {2002})},\ \Eprint
  {http://arxiv.org/abs/hep-ph/0111075} {arXiv:hep-ph/0111075 [hep-ph]}
  \BibitemShut {NoStop}%
\bibitem [{\citenamefont {Lee}(2005)}]{sQGP1}%
  \BibitemOpen
  \bibfield  {author} {\bibinfo {author} {\bibfnamefont {T.~D.}\ \bibnamefont
  {Lee}},\ }\href {\doibase 10.1016/j.nuclphysa.2004.11.003} {\bibfield
  {journal} {\bibinfo  {journal} {Nucl. Phys.}\ }\textbf {\bibinfo {volume}
  {A750}},\ \bibinfo {pages} {1} (\bibinfo {year} {2005})}\BibitemShut
  {NoStop}%
\bibitem [{\citenamefont {Gyulassy}\ and\ \citenamefont
  {McLerran}(2005)}]{sQGP2}%
  \BibitemOpen
  \bibfield  {author} {\bibinfo {author} {\bibfnamefont {M.}~\bibnamefont
  {Gyulassy}}\ and\ \bibinfo {author} {\bibfnamefont {L.}~\bibnamefont
  {McLerran}},\ }\href {\doibase 10.1016/j.nuclphysa.2004.10.034} {\bibfield
  {journal} {\bibinfo  {journal} {Nucl. Phys.}\ }\textbf {\bibinfo {volume}
  {A750}},\ \bibinfo {pages} {30} (\bibinfo {year} {2005})},\ \Eprint
  {http://arxiv.org/abs/nucl-th/0405013} {arXiv:nucl-th/0405013 [nucl-th]}
  \BibitemShut {NoStop}%
\bibitem [{\citenamefont {Shuryak}(2005)}]{sQGP3}%
  \BibitemOpen
  \bibfield  {author} {\bibinfo {author} {\bibfnamefont {E.~V.}\ \bibnamefont
  {Shuryak}},\ }\href {\doibase 10.1016/j.nuclphysa.2004.10.022} {\bibfield
  {journal} {\bibinfo  {journal} {Nucl. Phys.}\ }\textbf {\bibinfo {volume}
  {A750}},\ \bibinfo {pages} {64} (\bibinfo {year} {2005})},\ \Eprint
  {http://arxiv.org/abs/hep-ph/0405066} {arXiv:hep-ph/0405066 [hep-ph]}
  \BibitemShut {NoStop}%
\bibitem [{\citenamefont {Hirano}\ and\ \citenamefont
  {Gyulassy}(2006)}]{Hirano:2005wx}%
  \BibitemOpen
  \bibfield  {author} {\bibinfo {author} {\bibfnamefont {T.}~\bibnamefont
  {Hirano}}\ and\ \bibinfo {author} {\bibfnamefont {M.}~\bibnamefont
  {Gyulassy}},\ }\href {\doibase 10.1016/j.nuclphysa.2006.02.005} {\bibfield
  {journal} {\bibinfo  {journal} {Nucl. Phys.}\ }\textbf {\bibinfo {volume}
  {A769}},\ \bibinfo {pages} {71} (\bibinfo {year} {2006})},\ \Eprint
  {http://arxiv.org/abs/nucl-th/0506049} {arXiv:nucl-th/0506049 [nucl-th]}
  \BibitemShut {NoStop}%
\bibitem [{\citenamefont {Dusling}\ \emph {et~al.}(2016)\citenamefont
  {Dusling}, \citenamefont {Li},\ and\ \citenamefont
  {Schenke}}]{Dusling:2015gta}%
  \BibitemOpen
  \bibfield  {author} {\bibinfo {author} {\bibfnamefont {K.}~\bibnamefont
  {Dusling}}, \bibinfo {author} {\bibfnamefont {W.}~\bibnamefont {Li}}, \ and\
  \bibinfo {author} {\bibfnamefont {B.}~\bibnamefont {Schenke}},\ }\href
  {\doibase 10.1142/S0218301316300022} {\bibfield  {journal} {\bibinfo
  {journal} {Int. J. Mod. Phys.}\ }\textbf {\bibinfo {volume} {E25}},\ \bibinfo
  {pages} {1630002} (\bibinfo {year} {2016})},\ \Eprint
  {http://arxiv.org/abs/1509.07939} {arXiv:1509.07939 [nucl-ex]} \BibitemShut
  {NoStop}%
\bibitem [{\citenamefont {Adam}\ \emph {et~al.}(2017)\citenamefont {Adam} \emph
  {et~al.}}]{ALICE:2017jyt}%
  \BibitemOpen
  \bibfield  {author} {\bibinfo {author} {\bibfnamefont {J.}~\bibnamefont
  {Adam}} \emph {et~al.} (\bibinfo {collaboration} {ALICE}),\ }\href {\doibase
  10.1038/nphys4111} {\bibfield  {journal} {\bibinfo  {journal} {Nature Phys.}\
  }\textbf {\bibinfo {volume} {13}},\ \bibinfo {pages} {535} (\bibinfo {year}
  {2017})},\ \Eprint {http://arxiv.org/abs/1606.07424} {arXiv:1606.07424
  [nucl-ex]} \BibitemShut {NoStop}%
\bibitem [{\citenamefont {Rafelski}\ and\ \citenamefont
  {Muller}(1982)}]{Rafelski:1982pu}%
  \BibitemOpen
  \bibfield  {author} {\bibinfo {author} {\bibfnamefont {J.}~\bibnamefont
  {Rafelski}}\ and\ \bibinfo {author} {\bibfnamefont {B.}~\bibnamefont
  {Muller}},\ }\href {\doibase 10.1103/PhysRevLett.48.1066,
  10.1103/PhysRevLett.56.2334} {\bibfield  {journal} {\bibinfo  {journal}
  {Phys. Rev. Lett.}\ }\textbf {\bibinfo {volume} {48}},\ \bibinfo {pages}
  {1066} (\bibinfo {year} {1982})},\ \bibinfo {note} {[Erratum: Phys. Rev.
  Lett.56,2334(1986)]}\BibitemShut {NoStop}%
\bibitem [{\citenamefont {Koch}\ \emph {et~al.}(1983)\citenamefont {Koch},
  \citenamefont {Rafelski},\ and\ \citenamefont {Greiner}}]{Koch:1982ij}%
  \BibitemOpen
  \bibfield  {author} {\bibinfo {author} {\bibfnamefont {P.}~\bibnamefont
  {Koch}}, \bibinfo {author} {\bibfnamefont {J.}~\bibnamefont {Rafelski}}, \
  and\ \bibinfo {author} {\bibfnamefont {W.}~\bibnamefont {Greiner}},\ }\href
  {\doibase 10.1016/0370-2693(83)90411-2} {\bibfield  {journal} {\bibinfo
  {journal} {Phys. Lett.}\ }\textbf {\bibinfo {volume} {123B}},\ \bibinfo
  {pages} {151} (\bibinfo {year} {1983})}\BibitemShut {NoStop}%
\bibitem [{\citenamefont {Koch}\ \emph {et~al.}(1986)\citenamefont {Koch},
  \citenamefont {Muller},\ and\ \citenamefont {Rafelski}}]{Koch:1986ud}%
  \BibitemOpen
  \bibfield  {author} {\bibinfo {author} {\bibfnamefont {P.}~\bibnamefont
  {Koch}}, \bibinfo {author} {\bibfnamefont {B.}~\bibnamefont {Muller}}, \ and\
  \bibinfo {author} {\bibfnamefont {J.}~\bibnamefont {Rafelski}},\ }\href
  {\doibase 10.1016/0370-1573(86)90096-7} {\bibfield  {journal} {\bibinfo
  {journal} {Phys. Rept.}\ }\textbf {\bibinfo {volume} {142}},\ \bibinfo
  {pages} {167} (\bibinfo {year} {1986})}\BibitemShut {NoStop}%
\bibitem [{\citenamefont {Andersen}\ \emph {et~al.}(1999)\citenamefont
  {Andersen} \emph {et~al.}}]{Andersen:1999ym}%
  \BibitemOpen
  \bibfield  {author} {\bibinfo {author} {\bibfnamefont {E.}~\bibnamefont
  {Andersen}} \emph {et~al.} (\bibinfo {collaboration} {WA97}),\ }\href
  {\doibase 10.1016/S0370-2693(99)00140-9} {\bibfield  {journal} {\bibinfo
  {journal} {Phys. Lett.}\ }\textbf {\bibinfo {volume} {B449}},\ \bibinfo
  {pages} {401} (\bibinfo {year} {1999})}\BibitemShut {NoStop}%
\bibitem [{\citenamefont {Afanasiev}\ \emph {et~al.}(2002)\citenamefont
  {Afanasiev} \emph {et~al.}}]{Afanasiev:2002he}%
  \BibitemOpen
  \bibfield  {author} {\bibinfo {author} {\bibfnamefont {S.~V.}\ \bibnamefont
  {Afanasiev}} \emph {et~al.} (\bibinfo {collaboration} {NA49}),\ }\href
  {\doibase 10.1016/S0370-2693(02)01970-6} {\bibfield  {journal} {\bibinfo
  {journal} {Phys. Lett.}\ }\textbf {\bibinfo {volume} {B538}},\ \bibinfo
  {pages} {275} (\bibinfo {year} {2002})},\ \Eprint
  {http://arxiv.org/abs/hep-ex/0202037} {arXiv:hep-ex/0202037 [hep-ex]}
  \BibitemShut {NoStop}%
\bibitem [{\citenamefont {Antinori}\ \emph {et~al.}(2004)\citenamefont
  {Antinori} \emph {et~al.}}]{Antinori:2004ee}%
  \BibitemOpen
  \bibfield  {author} {\bibinfo {author} {\bibfnamefont {F.}~\bibnamefont
  {Antinori}} \emph {et~al.} (\bibinfo {collaboration} {NA57}),\ }\href
  {\doibase 10.1016/j.physletb.2004.05.025} {\bibfield  {journal} {\bibinfo
  {journal} {Phys. Lett.}\ }\textbf {\bibinfo {volume} {B595}},\ \bibinfo
  {pages} {68} (\bibinfo {year} {2004})},\ \Eprint
  {http://arxiv.org/abs/nucl-ex/0403022} {arXiv:nucl-ex/0403022 [nucl-ex]}
  \BibitemShut {NoStop}%
\bibitem [{\citenamefont {Abelev}\ \emph {et~al.}(2008)\citenamefont {Abelev}
  \emph {et~al.}}]{Abelev:2007xp}%
  \BibitemOpen
  \bibfield  {author} {\bibinfo {author} {\bibfnamefont {B.~I.}\ \bibnamefont
  {Abelev}} \emph {et~al.} (\bibinfo {collaboration} {STAR}),\ }\href {\doibase
  10.1103/PhysRevC.77.044908} {\bibfield  {journal} {\bibinfo  {journal} {Phys.
  Rev.}\ }\textbf {\bibinfo {volume} {C77}},\ \bibinfo {pages} {044908}
  (\bibinfo {year} {2008})},\ \Eprint {http://arxiv.org/abs/0705.2511}
  {arXiv:0705.2511 [nucl-ex]} \BibitemShut {NoStop}%
\bibitem [{\citenamefont {Abelev}\ \emph
  {et~al.}(2014{\natexlab{a}})\citenamefont {Abelev} \emph
  {et~al.}}]{ABELEV:2013zaa}%
  \BibitemOpen
  \bibfield  {author} {\bibinfo {author} {\bibfnamefont {B.~B.}\ \bibnamefont
  {Abelev}} \emph {et~al.} (\bibinfo {collaboration} {ALICE}),\ }\href
  {\doibase 10.1016/j.physletb.2014.05.052, 10.1016/j.physletb.2013.11.048}
  {\bibfield  {journal} {\bibinfo  {journal} {Phys. Lett.}\ }\textbf {\bibinfo
  {volume} {B728}},\ \bibinfo {pages} {216} (\bibinfo {year}
  {2014}{\natexlab{a}})},\ \bibinfo {note} {[Erratum: Phys.
  Lett.B734,409(2014)]},\ \Eprint {http://arxiv.org/abs/1307.5543}
  {arXiv:1307.5543 [nucl-ex]} \BibitemShut {NoStop}%
\bibitem [{\citenamefont {Sj{\"o}strand}\ \emph {et~al.}(2008)\citenamefont
  {Sj{\"o}strand}, \citenamefont {Mrenna},\ and\ \citenamefont
  {Skands}}]{Sjostrand:2007gs}%
  \BibitemOpen
  \bibfield  {author} {\bibinfo {author} {\bibfnamefont {T.}~\bibnamefont
  {Sj{\"o}strand}}, \bibinfo {author} {\bibfnamefont {S.}~\bibnamefont
  {Mrenna}}, \ and\ \bibinfo {author} {\bibfnamefont {P.~Z.}\ \bibnamefont
  {Skands}},\ }\href {\doibase 10.1016/j.cpc.2008.01.036} {\bibfield  {journal}
  {\bibinfo  {journal} {Comput. Phys. Commun.}\ }\textbf {\bibinfo {volume}
  {178}},\ \bibinfo {pages} {852} (\bibinfo {year} {2008})},\ \Eprint
  {http://arxiv.org/abs/0710.3820} {arXiv:0710.3820 [hep-ph]} \BibitemShut
  {NoStop}%
\bibitem [{\citenamefont {Andronic}\ \emph {et~al.}(2018)\citenamefont
  {Andronic}, \citenamefont {Braun-Munzinger}, \citenamefont {Redlich},\ and\
  \citenamefont {Stachel}}]{Andronic:2017pug}%
  \BibitemOpen
  \bibfield  {author} {\bibinfo {author} {\bibfnamefont {A.}~\bibnamefont
  {Andronic}}, \bibinfo {author} {\bibfnamefont {P.}~\bibnamefont
  {Braun-Munzinger}}, \bibinfo {author} {\bibfnamefont {K.}~\bibnamefont
  {Redlich}}, \ and\ \bibinfo {author} {\bibfnamefont {J.}~\bibnamefont
  {Stachel}},\ }\href {\doibase 10.1038/s41586-018-0491-6} {\bibfield
  {journal} {\bibinfo  {journal} {Nature}\ }\textbf {\bibinfo {volume} {561}},\
  \bibinfo {pages} {321} (\bibinfo {year} {2018})},\ \Eprint
  {http://arxiv.org/abs/1710.09425} {arXiv:1710.09425 [nucl-th]} \BibitemShut
  {NoStop}%
\bibitem [{\citenamefont {Werner}(2007)}]{Werner:2007bf}%
  \BibitemOpen
  \bibfield  {author} {\bibinfo {author} {\bibfnamefont {K.}~\bibnamefont
  {Werner}},\ }\href {\doibase 10.1103/PhysRevLett.98.152301} {\bibfield
  {journal} {\bibinfo  {journal} {Phys. Rev. Lett.}\ }\textbf {\bibinfo
  {volume} {98}},\ \bibinfo {pages} {152301} (\bibinfo {year} {2007})},\
  \Eprint {http://arxiv.org/abs/0704.1270} {arXiv:0704.1270 [nucl-th]}
  \BibitemShut {NoStop}%
\bibitem [{\citenamefont {Aichelin}\ and\ \citenamefont
  {Werner}(2009)}]{Aichelin:2008mi}%
  \BibitemOpen
  \bibfield  {author} {\bibinfo {author} {\bibfnamefont {J.}~\bibnamefont
  {Aichelin}}\ and\ \bibinfo {author} {\bibfnamefont {K.}~\bibnamefont
  {Werner}},\ }\href {\doibase 10.1103/PhysRevC.79.064907,
  10.1103/PhysRevC.81.029902} {\bibfield  {journal} {\bibinfo  {journal} {Phys.
  Rev.}\ }\textbf {\bibinfo {volume} {C79}},\ \bibinfo {pages} {064907}
  (\bibinfo {year} {2009})},\ \bibinfo {note} {[Erratum: Phys.
  Rev.C81,029902(2010)]},\ \Eprint {http://arxiv.org/abs/0810.4465}
  {arXiv:0810.4465 [nucl-th]} \BibitemShut {NoStop}%
\bibitem [{\citenamefont {Becattini}\ and\ \citenamefont
  {Manninen}(2009)}]{Becattini:2008ya}%
  \BibitemOpen
  \bibfield  {author} {\bibinfo {author} {\bibfnamefont {F.}~\bibnamefont
  {Becattini}}\ and\ \bibinfo {author} {\bibfnamefont {J.}~\bibnamefont
  {Manninen}},\ }\href {\doibase 10.1016/j.physletb.2009.01.066} {\bibfield
  {journal} {\bibinfo  {journal} {Phys. Lett.}\ }\textbf {\bibinfo {volume}
  {B673}},\ \bibinfo {pages} {19} (\bibinfo {year} {2009})},\ \Eprint
  {http://arxiv.org/abs/0811.3766} {arXiv:0811.3766 [nucl-th]} \BibitemShut
  {NoStop}%
\bibitem [{\citenamefont {Steinheimer}\ and\ \citenamefont
  {Bleicher}(2011)}]{Steinheimer:2011mp}%
  \BibitemOpen
  \bibfield  {author} {\bibinfo {author} {\bibfnamefont {J.}~\bibnamefont
  {Steinheimer}}\ and\ \bibinfo {author} {\bibfnamefont {M.}~\bibnamefont
  {Bleicher}},\ }\href {\doibase 10.1103/PhysRevC.84.024905} {\bibfield
  {journal} {\bibinfo  {journal} {Phys. Rev.}\ }\textbf {\bibinfo {volume}
  {C84}},\ \bibinfo {pages} {024905} (\bibinfo {year} {2011})},\ \Eprint
  {http://arxiv.org/abs/1104.3981} {arXiv:1104.3981 [hep-ph]} \BibitemShut
  {NoStop}%
\bibitem [{\citenamefont {Pierog}\ \emph {et~al.}(2015)\citenamefont {Pierog},
  \citenamefont {Karpenko}, \citenamefont {Katzy}, \citenamefont {Yatsenko},\
  and\ \citenamefont {Werner}}]{Pierog:2013ria}%
  \BibitemOpen
  \bibfield  {author} {\bibinfo {author} {\bibfnamefont {T.}~\bibnamefont
  {Pierog}}, \bibinfo {author} {\bibfnamefont {I.}~\bibnamefont {Karpenko}},
  \bibinfo {author} {\bibfnamefont {J.~M.}\ \bibnamefont {Katzy}}, \bibinfo
  {author} {\bibfnamefont {E.}~\bibnamefont {Yatsenko}}, \ and\ \bibinfo
  {author} {\bibfnamefont {K.}~\bibnamefont {Werner}},\ }\href {\doibase
  10.1103/PhysRevC.92.034906} {\bibfield  {journal} {\bibinfo  {journal} {Phys.
  Rev.}\ }\textbf {\bibinfo {volume} {C92}},\ \bibinfo {pages} {034906}
  (\bibinfo {year} {2015})},\ \Eprint {http://arxiv.org/abs/1306.0121}
  {arXiv:1306.0121 [hep-ph]} \BibitemShut {NoStop}%
\bibitem [{\citenamefont {Petrovici}\ \emph {et~al.}(2017)\citenamefont
  {Petrovici}, \citenamefont {Berceanu}, \citenamefont {Pop}, \citenamefont
  {Tarzila},\ and\ \citenamefont {Andrei}}]{Petrovici:2017izo}%
  \BibitemOpen
  \bibfield  {author} {\bibinfo {author} {\bibfnamefont {M.}~\bibnamefont
  {Petrovici}}, \bibinfo {author} {\bibfnamefont {I.}~\bibnamefont {Berceanu}},
  \bibinfo {author} {\bibfnamefont {A.}~\bibnamefont {Pop}}, \bibinfo {author}
  {\bibfnamefont {M.}~\bibnamefont {Tarzila}}, \ and\ \bibinfo {author}
  {\bibfnamefont {C.}~\bibnamefont {Andrei}},\ }\href {\doibase
  10.1103/PhysRevC.96.014908} {\bibfield  {journal} {\bibinfo  {journal} {Phys.
  Rev.}\ }\textbf {\bibinfo {volume} {C96}},\ \bibinfo {pages} {014908}
  (\bibinfo {year} {2017})},\ \Eprint {http://arxiv.org/abs/1703.05805}
  {arXiv:1703.05805 [nucl-th]} \BibitemShut {NoStop}%
\bibitem [{\citenamefont {Werner}\ \emph {et~al.}(2018)\citenamefont {Werner},
  \citenamefont {Knospe}, \citenamefont {Markert}, \citenamefont {Guiot},
  \citenamefont {Karpenko}, \citenamefont {Pierog}, \citenamefont {Sophys},
  \citenamefont {Stefaniak}, \citenamefont {Bleicher},\ and\ \citenamefont
  {Steinheimer}}]{Werner:2018SQM}%
  \BibitemOpen
  \bibfield  {author} {\bibinfo {author} {\bibfnamefont {K.}~\bibnamefont
  {Werner}}, \bibinfo {author} {\bibfnamefont {A.~G.}\ \bibnamefont {Knospe}},
  \bibinfo {author} {\bibfnamefont {C.}~\bibnamefont {Markert}}, \bibinfo
  {author} {\bibfnamefont {B.}~\bibnamefont {Guiot}}, \bibinfo {author}
  {\bibfnamefont {I.}~\bibnamefont {Karpenko}}, \bibinfo {author}
  {\bibfnamefont {T.}~\bibnamefont {Pierog}}, \bibinfo {author} {\bibfnamefont
  {G.}~\bibnamefont {Sophys}}, \bibinfo {author} {\bibfnamefont
  {M.}~\bibnamefont {Stefaniak}}, \bibinfo {author} {\bibfnamefont
  {M.}~\bibnamefont {Bleicher}}, \ and\ \bibinfo {author} {\bibfnamefont
  {J.}~\bibnamefont {Steinheimer}},\ }\href {\doibase
  10.1051/epjconf/201817109002} {\bibfield  {journal} {\bibinfo  {journal} {EPJ
  Web Conf.}\ }\textbf {\bibinfo {volume} {171}},\ \bibinfo {pages} {09002}
  (\bibinfo {year} {2018})}\BibitemShut {NoStop}%
\bibitem [{\citenamefont {Akamatsu}\ \emph {et~al.}(2018)\citenamefont
  {Akamatsu}, \citenamefont {Asakawa}, \citenamefont {Hirano}, \citenamefont
  {Kitazawa}, \citenamefont {Morita}, \citenamefont {Murase}, \citenamefont
  {Nara}, \citenamefont {Nonaka},\ and\ \citenamefont
  {Ohnishi}}]{Akamatsu:2018olk}%
  \BibitemOpen
  \bibfield  {author} {\bibinfo {author} {\bibfnamefont {Y.}~\bibnamefont
  {Akamatsu}}, \bibinfo {author} {\bibfnamefont {M.}~\bibnamefont {Asakawa}},
  \bibinfo {author} {\bibfnamefont {T.}~\bibnamefont {Hirano}}, \bibinfo
  {author} {\bibfnamefont {M.}~\bibnamefont {Kitazawa}}, \bibinfo {author}
  {\bibfnamefont {K.}~\bibnamefont {Morita}}, \bibinfo {author} {\bibfnamefont
  {K.}~\bibnamefont {Murase}}, \bibinfo {author} {\bibfnamefont
  {Y.}~\bibnamefont {Nara}}, \bibinfo {author} {\bibfnamefont {C.}~\bibnamefont
  {Nonaka}}, \ and\ \bibinfo {author} {\bibfnamefont {A.}~\bibnamefont
  {Ohnishi}},\ }\href {\doibase 10.1103/PhysRevC.98.024909} {\bibfield
  {journal} {\bibinfo  {journal} {Phys. Rev.}\ }\textbf {\bibinfo {volume}
  {C98}},\ \bibinfo {pages} {024909} (\bibinfo {year} {2018})},\ \Eprint
  {http://arxiv.org/abs/1805.09024} {arXiv:1805.09024 [nucl-th]} \BibitemShut
  {NoStop}%
\bibitem [{\citenamefont {Bozek}(2005)}]{Bozek:2005eu}%
  \BibitemOpen
  \bibfield  {author} {\bibinfo {author} {\bibfnamefont {P.}~\bibnamefont
  {Bozek}},\ }\href@noop {} {\bibfield  {journal} {\bibinfo  {journal} {Acta
  Phys. Polon.}\ }\textbf {\bibinfo {volume} {B36}},\ \bibinfo {pages} {3071}
  (\bibinfo {year} {2005})},\ \Eprint {http://arxiv.org/abs/nucl-th/0506037}
  {arXiv:nucl-th/0506037 [nucl-th]} \BibitemShut {NoStop}%
\bibitem [{\citenamefont {Bozek}(2009)}]{Bozek:2008zw}%
  \BibitemOpen
  \bibfield  {author} {\bibinfo {author} {\bibfnamefont {P.}~\bibnamefont
  {Bozek}},\ }\href {\doibase 10.1103/PhysRevC.79.054901} {\bibfield  {journal}
  {\bibinfo  {journal} {Phys. Rev.}\ }\textbf {\bibinfo {volume} {C79}},\
  \bibinfo {pages} {054901} (\bibinfo {year} {2009})},\ \Eprint
  {http://arxiv.org/abs/0811.1918} {arXiv:0811.1918 [nucl-th]} \BibitemShut
  {NoStop}%
\bibitem [{\citenamefont {Andersson}\ \emph {et~al.}(1987)\citenamefont
  {Andersson}, \citenamefont {Gustafson},\ and\ \citenamefont
  {Nilsson-Almqvist}}]{ANDERSSON1987289}%
  \BibitemOpen
  \bibfield  {author} {\bibinfo {author} {\bibfnamefont {B.}~\bibnamefont
  {Andersson}}, \bibinfo {author} {\bibfnamefont {G.}~\bibnamefont
  {Gustafson}}, \ and\ \bibinfo {author} {\bibfnamefont {B.}~\bibnamefont
  {Nilsson-Almqvist}},\ }\href {\doibase
  https://doi.org/10.1016/0550-3213(87)90257-4} {\bibfield  {journal} {\bibinfo
   {journal} {Nuclear Physics B}\ }\textbf {\bibinfo {volume} {281}},\ \bibinfo
  {pages} {289 } (\bibinfo {year} {1987})}\BibitemShut {NoStop}%
\bibitem [{\citenamefont {Bierlich}\ \emph {et~al.}(2016)\citenamefont
  {Bierlich}, \citenamefont {Gustafson},\ and\ \citenamefont
  {L{\"o}nnblad}}]{Bierlich:2016smv}%
  \BibitemOpen
  \bibfield  {author} {\bibinfo {author} {\bibfnamefont {C.}~\bibnamefont
  {Bierlich}}, \bibinfo {author} {\bibfnamefont {G.}~\bibnamefont {Gustafson}},
  \ and\ \bibinfo {author} {\bibfnamefont {L.}~\bibnamefont {L{\"o}nnblad}},\
  }\href {\doibase 10.1007/JHEP10(2016)139} {\bibfield  {journal} {\bibinfo
  {journal} {JHEP}\ }\textbf {\bibinfo {volume} {10}},\ \bibinfo {pages} {139}
  (\bibinfo {year} {2016})},\ \Eprint {http://arxiv.org/abs/1607.04434}
  {arXiv:1607.04434 [hep-ph]} \BibitemShut {NoStop}%
\bibitem [{\citenamefont {Okai}\ \emph {et~al.}(2017)\citenamefont {Okai},
  \citenamefont {Kawaguchi}, \citenamefont {Tachibana},\ and\ \citenamefont
  {Hirano}}]{Okai:2017ofp}%
  \BibitemOpen
  \bibfield  {author} {\bibinfo {author} {\bibfnamefont {M.}~\bibnamefont
  {Okai}}, \bibinfo {author} {\bibfnamefont {K.}~\bibnamefont {Kawaguchi}},
  \bibinfo {author} {\bibfnamefont {Y.}~\bibnamefont {Tachibana}}, \ and\
  \bibinfo {author} {\bibfnamefont {T.}~\bibnamefont {Hirano}},\ }\href
  {\doibase 10.1103/PhysRevC.95.054914} {\bibfield  {journal} {\bibinfo
  {journal} {Phys. Rev.}\ }\textbf {\bibinfo {volume} {C95}},\ \bibinfo {pages}
  {054914} (\bibinfo {year} {2017})},\ \Eprint
  {http://arxiv.org/abs/1702.07541} {arXiv:1702.07541 [nucl-th]} \BibitemShut
  {NoStop}%
\bibitem [{\citenamefont {Shen}\ and\ \citenamefont
  {Schenke}(2018)}]{Shen:2017bsr}%
  \BibitemOpen
  \bibfield  {author} {\bibinfo {author} {\bibfnamefont {C.}~\bibnamefont
  {Shen}}\ and\ \bibinfo {author} {\bibfnamefont {B.}~\bibnamefont {Schenke}},\
  }\href {\doibase 10.1103/PhysRevC.97.024907} {\bibfield  {journal} {\bibinfo
  {journal} {Phys. Rev.}\ }\textbf {\bibinfo {volume} {C97}},\ \bibinfo {pages}
  {024907} (\bibinfo {year} {2018})},\ \Eprint
  {http://arxiv.org/abs/1710.00881} {arXiv:1710.00881 [nucl-th]} \BibitemShut
  {NoStop}%
\bibitem [{\citenamefont {Borsanyi}\ \emph {et~al.}(2014)\citenamefont
  {Borsanyi}, \citenamefont {Fodor}, \citenamefont {Hoelbling}, \citenamefont
  {Katz}, \citenamefont {Krieg},\ and\ \citenamefont
  {Szabo}}]{Borsanyi:2013cga}%
  \BibitemOpen
  \bibfield  {author} {\bibinfo {author} {\bibfnamefont {S.}~\bibnamefont
  {Borsanyi}}, \bibinfo {author} {\bibfnamefont {Z.}~\bibnamefont {Fodor}},
  \bibinfo {author} {\bibfnamefont {C.}~\bibnamefont {Hoelbling}}, \bibinfo
  {author} {\bibfnamefont {S.~D.}\ \bibnamefont {Katz}}, \bibinfo {author}
  {\bibfnamefont {S.}~\bibnamefont {Krieg}}, \ and\ \bibinfo {author}
  {\bibfnamefont {K.~K.}\ \bibnamefont {Szabo}},\ }\href@noop {} {\bibfield
  {journal} {\bibinfo  {journal} {Phys. Lett.}\ }\textbf {\bibinfo {volume}
  {B730}},\ \bibinfo {pages} {155} (\bibinfo {year} {2014})},\ \Eprint
  {http://arxiv.org/abs/1312.2193} {arXiv:1312.2193 [hep-lat]} \BibitemShut
  {NoStop}%
\bibitem [{\citenamefont {Cooper}\ and\ \citenamefont
  {Frye}(1974)}]{Cooper:1974mv}%
  \BibitemOpen
  \bibfield  {author} {\bibinfo {author} {\bibfnamefont {F.}~\bibnamefont
  {Cooper}}\ and\ \bibinfo {author} {\bibfnamefont {G.}~\bibnamefont {Frye}},\
  }\href {\doibase 10.1103/PhysRevD.10.186} {\bibfield  {journal} {\bibinfo
  {journal} {Phys. Rev.}\ }\textbf {\bibinfo {volume} {D10}},\ \bibinfo {pages}
  {186} (\bibinfo {year} {1974})}\BibitemShut {NoStop}%
\bibitem [{\citenamefont {Takeuchi}\ \emph {et~al.}(2015)\citenamefont
  {Takeuchi}, \citenamefont {Murase}, \citenamefont {Hirano}, \citenamefont
  {Huovinen},\ and\ \citenamefont {Nara}}]{Takeuchi:2015ana}%
  \BibitemOpen
  \bibfield  {author} {\bibinfo {author} {\bibfnamefont {S.}~\bibnamefont
  {Takeuchi}}, \bibinfo {author} {\bibfnamefont {K.}~\bibnamefont {Murase}},
  \bibinfo {author} {\bibfnamefont {T.}~\bibnamefont {Hirano}}, \bibinfo
  {author} {\bibfnamefont {P.}~\bibnamefont {Huovinen}}, \ and\ \bibinfo
  {author} {\bibfnamefont {Y.}~\bibnamefont {Nara}},\ }\href {\doibase
  10.1103/PhysRevC.92.044907} {\bibfield  {journal} {\bibinfo  {journal} {Phys.
  Rev.}\ }\textbf {\bibinfo {volume} {C92}},\ \bibinfo {pages} {044907}
  (\bibinfo {year} {2015})},\ \Eprint {http://arxiv.org/abs/1505.05961}
  {arXiv:1505.05961 [nucl-th]} \BibitemShut {NoStop}%
\bibitem [{\citenamefont {Abelev}\ \emph
  {et~al.}(2014{\natexlab{b}})\citenamefont {Abelev} \emph
  {et~al.}}]{Abelev:2013haa}%
  \BibitemOpen
  \bibfield  {author} {\bibinfo {author} {\bibfnamefont {B.~B.}\ \bibnamefont
  {Abelev}} \emph {et~al.} (\bibinfo {collaboration} {ALICE}),\ }\href
  {\doibase 10.1016/j.physletb.2013.11.020} {\bibfield  {journal} {\bibinfo
  {journal} {Phys. Lett.}\ }\textbf {\bibinfo {volume} {B728}},\ \bibinfo
  {pages} {25} (\bibinfo {year} {2014}{\natexlab{b}})},\ \Eprint
  {http://arxiv.org/abs/1307.6796} {arXiv:1307.6796 [nucl-ex]} \BibitemShut
  {NoStop}%
\bibitem [{\citenamefont {Adam}\ \emph {et~al.}(2016)\citenamefont {Adam} \emph
  {et~al.}}]{Adam:2015vsf}%
  \BibitemOpen
  \bibfield  {author} {\bibinfo {author} {\bibfnamefont {J.}~\bibnamefont
  {Adam}} \emph {et~al.} (\bibinfo {collaboration} {ALICE}),\ }\href {\doibase
  10.1016/j.physletb.2016.05.027} {\bibfield  {journal} {\bibinfo  {journal}
  {Phys. Lett.}\ }\textbf {\bibinfo {volume} {B758}},\ \bibinfo {pages} {389}
  (\bibinfo {year} {2016})},\ \Eprint {http://arxiv.org/abs/1512.07227}
  {arXiv:1512.07227 [nucl-ex]} \BibitemShut {NoStop}%
\bibitem [{\citenamefont {Kanakubo}\ \emph {et~al.}(shed)\citenamefont
  {Kanakubo}, \citenamefont {Okai}, \citenamefont {Tachibana},\ and\
  \citenamefont {Hirano}}]{KanakuboFull}%
  \BibitemOpen
  \bibfield  {author} {\bibinfo {author} {\bibfnamefont {Y.}~\bibnamefont
  {Kanakubo}}, \bibinfo {author} {\bibfnamefont {M.}~\bibnamefont {Okai}},
  \bibinfo {author} {\bibfnamefont {Y.}~\bibnamefont {Tachibana}}, \ and\
  \bibinfo {author} {\bibfnamefont {T.}~\bibnamefont {Hirano}},\ }\href@noop {}
  {\bibfield  {journal} {\bibinfo  {journal} {in preparation}\ } (\bibinfo
  {year} {unpublished})}\BibitemShut {NoStop}%
\end{thebibliography}%

\end{document}